\documentstyle[12pt,blois,psfig]{article}

\begin{document}

\heading{HYBRID GALAXY FORMATION}

\author{A. J. Benson$^1$, S. Cole$^1$, C. S. Frenk$^1$, C. M. Baugh$^1$ and C. G. Lacey$^2$} {$^1$ Department of Physics, University of Durham, Durham, UK.} {$^2$ Theoretical Astrophysics Center, Copenhagen, Denmark.}

\begin{bloisabstract}
We model the two-point correlation function of galaxies in a cold dark matter Universe by combining two powerful theoretical tools --- dissipationless N-body simulations of dark matter clustering (specifically the GIF simulations carried out by MPIA and the Virgo Consortium) and semi-analytic modelling of galaxy formation. We construct catalogues of galaxies containing a wide range of information for each galaxy, including magnitudes in various bands, star formation rates, disk and bulge sizes (from the semi-analytic model described by Cole et al \cite{coleetal98}) and, importantly, spatial positions and peculiar velocities from the N-body simulation. We then use this information to study the clustering properties of galaxies.
\end{bloisabstract}

\section{Catalogue construction}

Whilst semi-analytic models have proven useful in modelling and predicting numerous properties of the galaxy population (Kauffmann, White \& Guiderdoni \cite{kwg}; Cole et al \cite{coleetal94}; Somerville \& Primack \cite{somervilleprimack} they provide only limited information on the spatial distribution of galaxies, and so it has been difficult to study in detail the clustering properties of galaxies using these models. Recently, semi-analytic models have been combined with N-body simulations to provide the required spatial information (Kauffmann, Nusser \& Steinmetz \cite{kns}; Governato et al \cite{fabio}; Kauffmann et al \cite{gketal}). We have used a similar technique to study the clustering of galaxies in CDM universes within well constrained semi-analytic models. The main difference between our technique and that of Kauffmann et al \cite{gketal} is that whilst they extract the merging history of each dark matter halo directly from the simulation, we construct this history using the extended Press-Schechter theory. We find that this gives the same statistical results as the Kauffmann et al method and allows us to resolve merger trees to much smaller masses.

To construct a catalogue of galaxies containing spatial information we use the following procedure: (i) take the output from a dissipationless N-body simulation of dark matter and use a group finding algorithm (here we use the friends-of-friends algorithm with the standard linking length of $b=0.2$) to locate bound, virialised haloes of dark matter of 10 or more particles (such groups have been shown to be stable by Kauffmann et al \cite{gketal}); (ii) determine the mass of each group, the position and velocity of its centre of mass and the positions and velocities of randomly selected particles within the group; (iii) for each group use a semi-analytic model of galaxy formation constrained to match the local B and K-band luminosity functions to determine the population of galaxies living within the dark matter halo; (iv) attach the central galaxy of the halo to the centre of mass of the group and attach any satellite galaxies to randomly selected particles within the halo so that galaxies trace mass within a given dark matter halo (which may not be exactly true in reality because of processes such as dynamical friction).

\begin{figure}
\centering
\begin{tabular}{cc}
\psfig{file=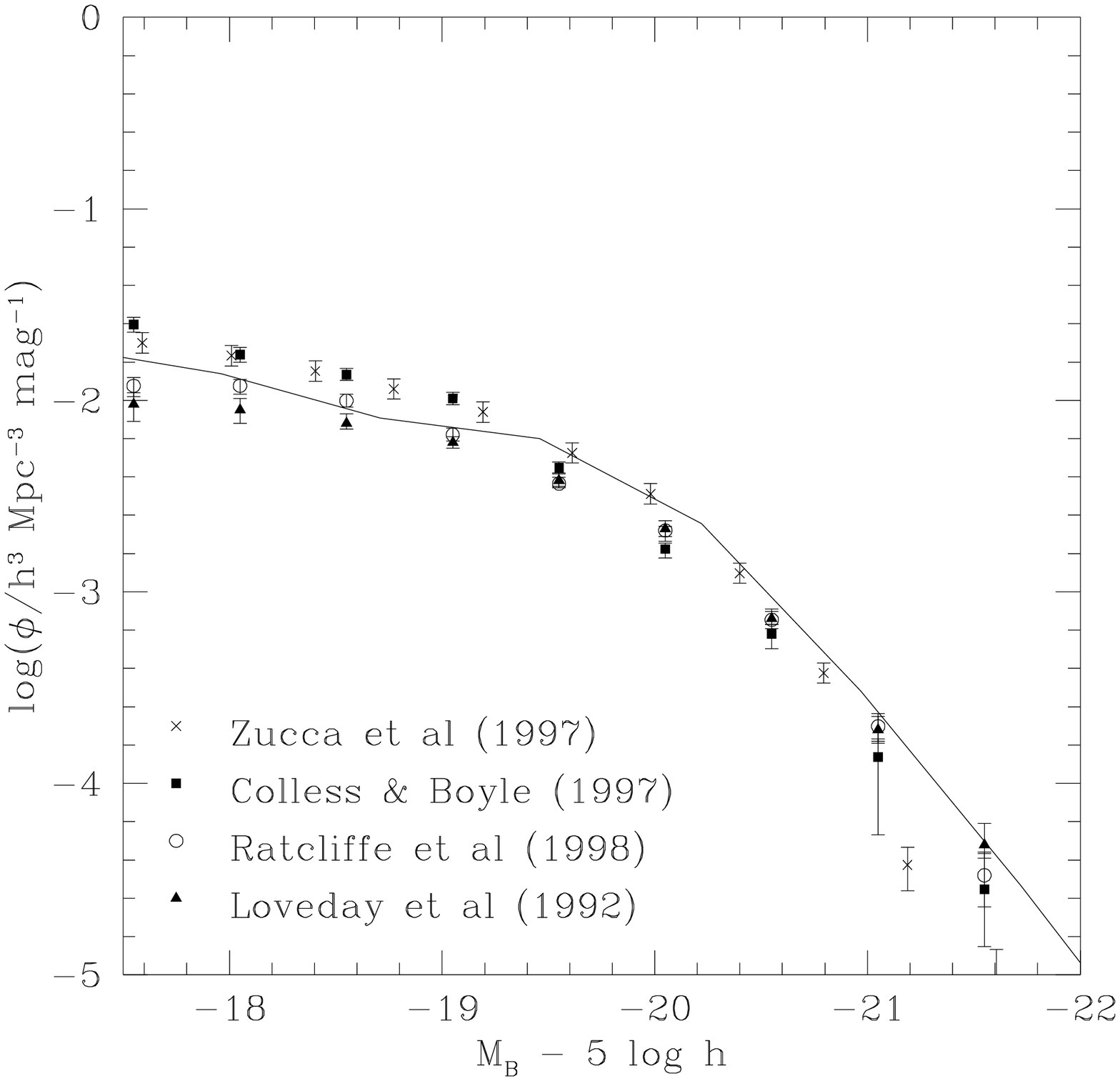,width=80mm} & \psfig{file=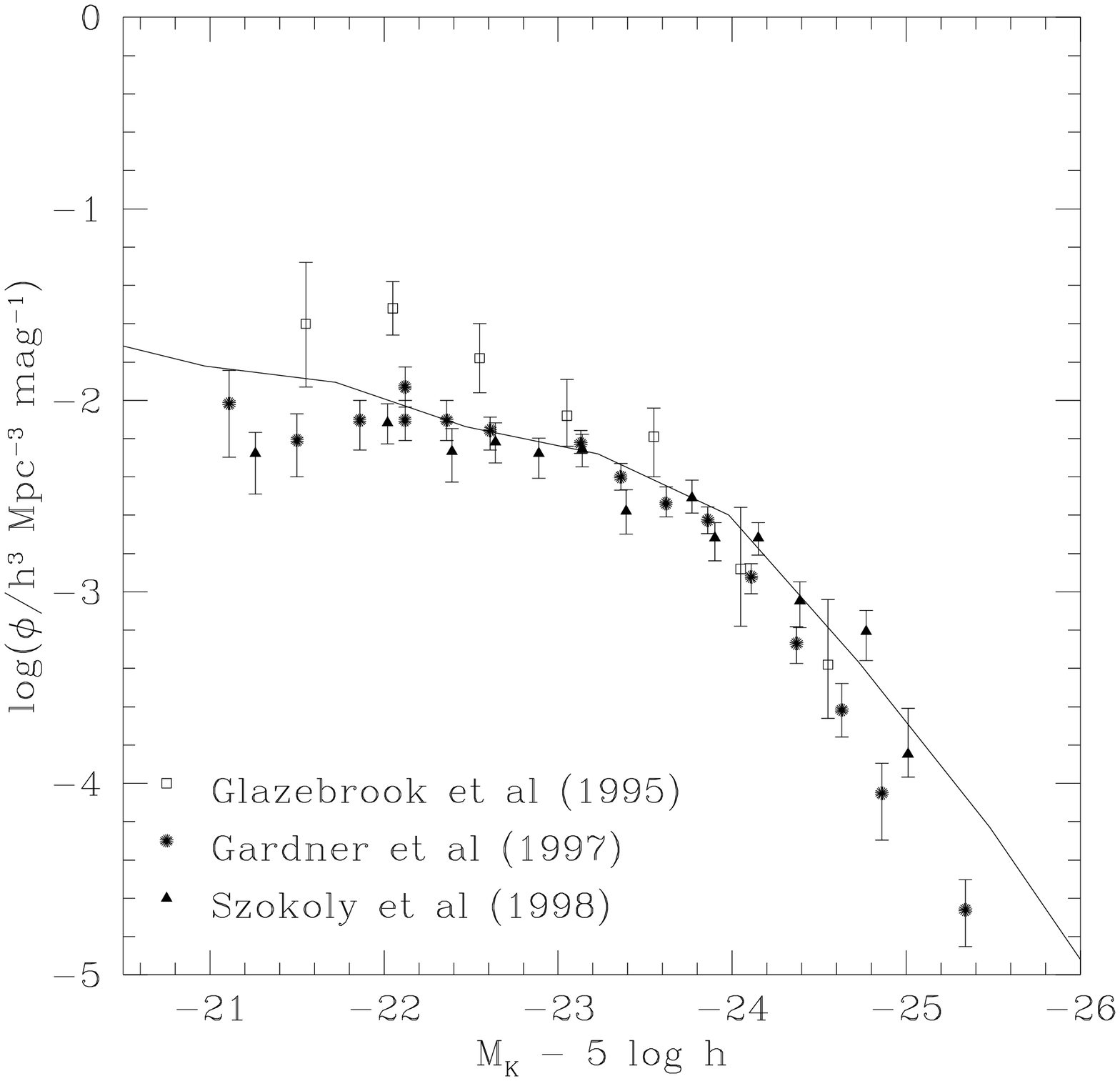,width=80mm}
\end{tabular}
\caption{The local B and K-band luminosity functions from our model compared to various observational determinations. Note the good agreement between model (solid line) and observations (symbols) at the bright end. Our galaxy catalogues become incomplete at the faint end of the luminosity functions (faintwards of $M_{\mathrm B} - 5 \log h \approx -17.5$ in the B-band and $M_{\mathrm K} - 5 \log h \approx -20.5$ in the K-band) due to the limited resolution of the N-body simulation. Our catalogues are constructed using only galaxies for which a complete sample is available within the model.}
\label{fig1}
\end{figure}

\begin{figure}
\centering
\hspace{0.in}\psfig{file=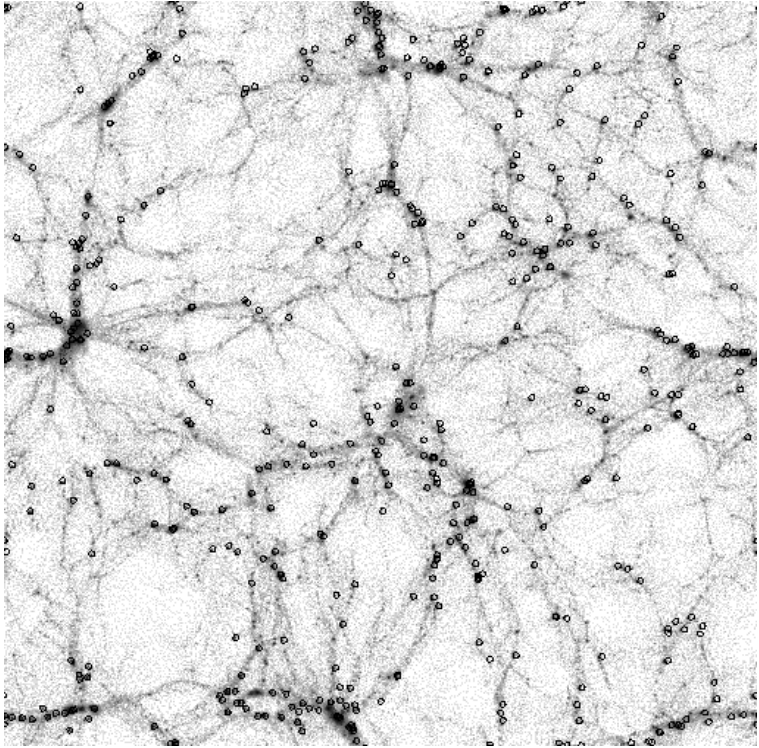,width=100mm}
\caption{A slice through a $\Lambda$CDM N-body simulation. The volume shown is 141 $\times$ 141 $\times$ 8 $h^{-1}$ Mpc. Dark matter is shown by the greyscale, with the darker areas being the most dense. Overlaid are the positions of all galaxies brighter than $M_{\mathrm B} - 5 \log h = -19.5$ indicated by circles.}
\label{fig2}
\end{figure}

This results in a galaxy catalogue which can be analysed to determine the clustering properties of galaxies of any given luminosity, morphology, colour and so on. In fact by using this technique it is possible to produce catalogues of galaxies complete with spatial information (or alternatively redshifts, and angular coordinates) with any observationally motivated selection criteria. Furthermore, by identifying dark matter halos on the past lightcone of an observer a full kock galaxy redshift survey can be constructed. The luminosity functions determined from the galaxy catalogue are shown in Figure \ref{fig1}. Although they become incomplete at the faint end because of the limited resolution of the N-body simulation there is good agreement between the bright ends and the observed luminosity functions. Since we only consider the clustering of galaxies for which our catalogue is complete the resolution limit is not important. We find however that an accurate match to the bright end of the luminosity function is important as it strongly constrains the resulting two-point correlation function. An example of the information produced by our model is given in Figure \ref{fig2}, which shows a slice through a $\Lambda$CDM N-body simulation upon which the positions of galaxies brighter than $M_{\mathrm B} - 5 \log h = -19.5$ (we define the Hubble constant to be ${\mathrm H}_0 = 100 h$ km s$^{-1}$ Mpc$^{-1}$) have been overlaid as circles. It can be seen quite clearly that the galaxies trace the mass to some extent. To determine exactly how well galaxies trace the underlying dark matter we estimate the two-point correlation function of these galaxies.

\section{The two-point correlation function}

\begin{figure}[t]
\centering
\hspace{0.in}\psfig{file=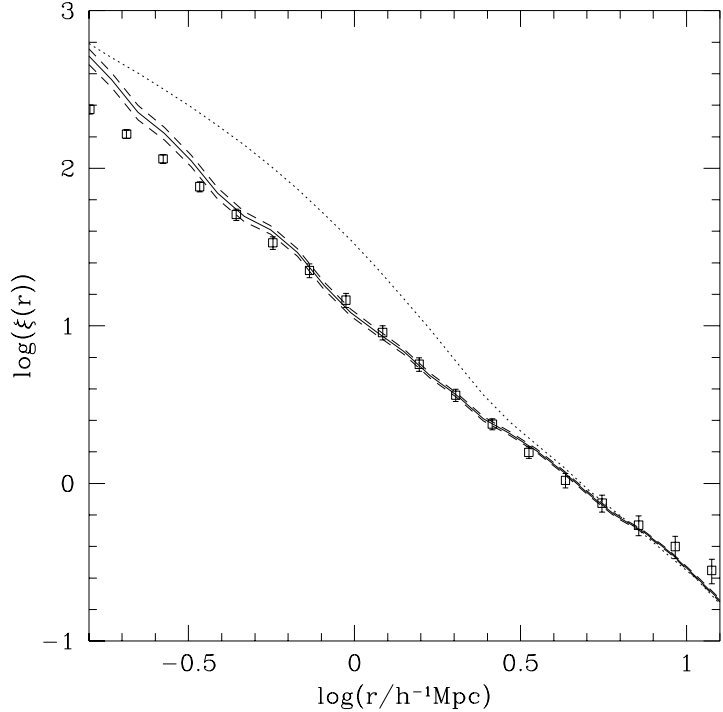,width=100mm}
\caption{The correlation function of galaxies brighter than $M_{\mathrm B} - 5 \log h = -19.5$ in a $\Lambda$CDM cosmology. Points with errorbars show the APM galaxy correlation function of Baugh \cite{APM}. The dotted line shows the dark matter correlation function whilst the solid line shows the correlation function of galaxies in our model (the dashed lines to either side indicate the Poisson errors on this correlation function).}
\label{fig3}
\end{figure}

Shown in Figure \ref{fig3} is the two-point correlation function of galaxies brighter than $M_{\mathrm B} - 5 \log h = -19.5$ in our $\Lambda$CDM model (which has $\Omega _0 = 0.3$, $\Lambda = 0.7$, $h = 0.7$ and $\sigma _8 = 0.9$). This is compared to the correlation function of the underlying dark matter and to the observationally determined correlation function of galaxies in the APM survey from Baugh \cite{APM}. Note that all of the curves shown here are real space correlation functions. The galaxy correlation function shows many interesting properties. Firstly it is biased with respect to the mass correlation function, and furthermore this bias is scale dependent. On small scales there is in fact an antibias, which is exactly what is needed to reconcile the theory with the observed galaxy correlation function. The observed and model galaxy correlation functions agree over a wide range of scales, both showing approximate power law behaviour. The scale-dependant bias seen in these models arises from a complex interplay of effects. On large scales the bias is due to the intrinsic bias of dark matter halos in a CDM universe as described by Mo \& White \cite{mowhite}. On smaller scales the bias is controlled by the way halos are populated with galaxies, specifically the variations in number of galaxies per halo for halos of a given mass. The Lagrangian radius exclusion of halos also affects the small scale bias. These issues are explored in greater detail by Benson et al \cite{meetal}, who also study $\Omega _0 = 1$ models and find that such models fail to reproduce the observed clustering of galaxies.

\acknowledgements{}

AJB's attendance at the X$^{\mathrm th}$ Rencontres de Blois was funded in part by a grant from the European Commision.

% References listed in alphabetical order ...

%
\vfill

\begin{bloisbib}
\bibitem{APM} Baugh C. M., 1996, \mnras {280} {267}
\bibitem{meetal} Benson A. J., Cole S., Frenk C. S., Baugh C. M., Lacey C. G., 1998, {\it in preparation}
\bibitem{coleetal94} Cole S., Lacey C. G., Baugh C. M., Frenk C. S., 1994, \mnras {271} {781}
\bibitem{coleetal98} Cole S., Lacey C. G., Baugh C. M., Frenk C. S., 1998, {\it in preparation}
\bibitem{colless} Colless M., Boyle B., 1997, {\it astro-ph/9710268}
\bibitem{gardner} Gardner J. P., Sharples R. M., Frenk C. S., Carrasco B. E., 1997, \apj {480} {L99}
\bibitem{glazebrook} Glazebrook K. et al, 1995, {\it astro-ph/9503116}
\bibitem{fabio} Governato F. et al., 1998, \nat {392} {359}
\bibitem{kwg} Kauffmann G., White S. D. M., Guiderdoni B., 1993, \mnras {264} {201}
\bibitem{kns} Kauffmann G., Nusser A., Steinmetz M., 1997, \mnras {286} {795}
\bibitem{gketal} Kauffmann G., Colberg J. M., Diaferio A., White S. D. M., 1998, {\it astro-ph/9805283}
\bibitem{loveday} Loveday J., Peterson B. A., Efstathiou G., Maddox S. J., 1992, \apj {390} {338}
\bibitem{mowhite} Mo H. J., White S. D. M., 1996, \mnras {282} {347}
\bibitem{arat} Ratcliffe A., Shanks T., Parker Q. A., Fong R., 1998, \mnras {296} {173}
\bibitem{somervilleprimack} Somerville R., Primack J., 1998, {\it astro-ph/9802268}
\bibitem{szokoly} Szokoly G.P., Subbarao M.U., Conolly A.J., Mobasher B., 1998, {\it astro-ph/9801132}
\bibitem{zucca} Zucca E. et al, 1997, \aa {326} {477}
\end{bloisbib}
\end{document}